\documentclass[12pt,english,floatfix,superscriptaddress,aps,prd,preprint,showkeys,nofootinbib]{revtex4}
\usepackage{amsmath}
\usepackage{amssymb}
\usepackage{amsbsy}
\usepackage{amsfonts}
\usepackage{amsopn}
\usepackage{amstext}
\usepackage{graphicx}
\usepackage[english]{babel}
\usepackage{color}
\usepackage{slashed}
\usepackage{esint}
\usepackage[dvips]{epsfig}
\usepackage[dvips]{graphicx}
\usepackage{float}
\usepackage{units}
\usepackage{esint}
\usepackage{textcomp}
\usepackage{wasysym}
\usepackage{hyperref}
\usepackage{slashed}
\usepackage{tikz}
\usetikzlibrary{decorations.pathmorphing,arrows.meta}

\newcommand{\dd}{\mathrm{d}}
\newcommand{\ii}{\mathrm{i}}

\newcommand{\eps}{\epsilon}

\newif\ifshowchanges
\showchangestrue

\begin{document}

\title{One-loop contributions in Lorentz-violating scalar QED with nonminimal coupling}

\author{F. M. Belchior}
\email{fernandobelcks8@gmail.com}
\affiliation{Departamento de Física, Universidade Federal da Paraíba, Centro de Ciências Exatas e da Natureza, 58051-970, João Pessoa, Paraíba, Brazil}

\author{J. R. Nascimento}
\email{jroberto@fisica.ufpb.br}
\affiliation{Departamento de Física, Universidade Federal da Paraíba, Centro de Ciências Exatas e da Natureza, 58051-970, João Pessoa, Paraíba, Brazil}

\author{A. Yu. Petrov}
\email{petrov@fisica.ufpb.br}
\affiliation{Departamento de Física, Universidade Federal da Paraíba, Centro de Ciências Exatas e da Natureza, 58051-970, João Pessoa, Paraíba, Brazil}

 \author{P. Porfirio}
\email{pporfirio@fisica.ufpb.br}
\affiliation{Departamento de Física, Universidade Federal da Paraíba, Centro de Ciências Exatas e da Natureza, 58051-970, João Pessoa, Paraíba, Brazil}

\begin{abstract}
In the context of pertubative aspects of Lorentz-violating theories, we study the divergent one-loop two-point functions of a Lorentz-violating (LV) extension of scalar QED containing a dimension-five CPT-odd nonminimal coupling. The model is defined through a generalized covariant derivative involving a constant background vector and the dual electromagnetic field strength. After expanding the action, we derive the full set of vertices relevant for the photon and scalar self-energies, including the mixed minimal-nonminimal seagull interaction diagram required by the gauge invariance. Using dimensional regularization, we compute the divergent parts of the vacuum polarization and scalar self-energy. The complete photon two-point function is shown to be transverse since the mass-dependent pieces generated by the mixed and purely nonminimal bubble graphs cancel against the corresponding seagull graphs, leaving higher-derivative CPT-odd and CPT-even LV gauge counterterms. The scalar self-energy generates LV kinetic and higher-derivative counterterms. The calculation confirms that gauge invariance is preserved, while the nonminimal interaction should be understood within an effective-field-theory expansion.

\end{abstract}
\keywords{Lorentz violation; scalar QED; nonminimal coupling; vacuum polarization; scalar self-energy; dimensional regularization}

\maketitle

\section{Introduction}
As a new tool to probe the physics at the Planck scale, where quantum gravity effects are expected to be observed, it is believed that the Lorentz symmetry might not be exact at such a scale. In this regard, models have been proposed to examine possible departures from  Lorentz symmetry, and this possibility has already been tested through various experiments, see \cite{Kostelecky:2011gq} and references therein. Studies on perturbative aspects of several LV models have become an important line of investigation in recent years. There are works related to both CPT-odd and CPT-even sectors of the LV Standard Model extension (LV SME)  which has been originally developed in \cite{Colladay:1996iz,Colladay:1998fq}. A noteworthy result was obtained in Ref. \cite{Jackiw:1999yp}, where the authors succeeded in generating  the Carroll-Field-Jackiw (CFJ) \cite{Carroll:1989vb} term by means of one-loop quantum correction in CPT-odd LV spinor QED. Various issues related both to classical and quantum aspects of the spinor QED have been studied in numerous works, see e.g. \cite{Chung:1998jv,Chung:1999pt,Kostelecky:2000mm,Kostelecky:2001jc,Kostelecky:2002ue,Cambiaso:2012vb,Ding:2016lwt} and many other papers. At the same time, besides the fermionic QED, another point of interest is to examine perturbative aspects of scalar QED, where the Higgs sector of the SME becomes relevant. Overall, to accomplish LV modifications, one can consider that vector and tensor background fields are coupled to the physical fields, leading to explicit Lorentz symmetry violation. For the sake of illustration, let us take as an example the following Lagrangian density for an extended LV scalar QED known to be an ingredient of the LV SME (cf. \cite{Colladay:1996iz,Colladay:1998fq}):
\begin{align}\label{SMEmodel}
\mathcal{L}=&-\frac{1}{4} F_{\mu\nu}F^{\mu\nu}+(\eta^{\mu\nu}+\kappa^{\mu\nu})(D_\mu\phi)^{\ast}(D_\nu\phi)-\frac{i}{2}(\phi^{\ast}k^\mu D_\mu\phi-\phi\, k^\mu D_\mu\phi^{\ast})\nonumber\\&-m^2\phi^{\ast}\phi-\frac{\lambda_\phi}{4}(\phi^{\ast}\phi)^2,
\end{align}
where $D_\mu=\partial_\mu+ieA_\mu$ represents the usual covariant derivative and $\eta^{\mu\nu}=diag(1,-1,-1,-1)$ is the Minkowski metric, while the vector $k^\mu$ and the tensor $\kappa^{\mu\nu}$ are constants generating the LV effects. It is important to mention that the tensor $\kappa^{\mu\nu}$ is required to be symmetric and traceless. Moreover, it should be emphasized that both the vector $k^\mu$ and the tensor $\kappa^{\mu\nu}$ implement the Lorentz symmetry violation only in the particle reference frame, implying the independence of the space-time position. Consequently, the translational invariance is ensured, leading to the conservation of momentum and energy as is physically desired. 
The coefficient $k^\mu$ in Eq.~\eqref{SMEmodel} is not an independent source
of observable Lorentz violation for one charged scalar.  Let
$q_\mu=\frac12(h^{-1})_{\mu\nu}k^\nu$. By completing the square, we arrive at
\begin{align}
h^{\mu\nu}(D_\mu\phi)^\ast(D_\nu\phi)-\frac{i}{2}k^\mu\!\left(\phi^\ast D_\mu\phi-\phi(D_\mu\phi)^\ast\right)={}&h^{\mu\nu}\big[(D_\mu+iq_\mu)\phi\big]^\ast
\big[(D_\nu+iq_\nu)\phi\big]\nonumber\\&-\frac14 k^\mu(h^{-1})_{\mu\nu}k^\nu\,\phi^\ast\phi ,
\end{align}
where the first term is understood as the modulus squared of
$(D_\mu+iq_\mu)\phi$.  Equivalently, the phase redefinition
$\phi(x)=e^{-iq\cdot x}\chi(x)$ removes $k^\mu$ and replaces the mass by
\begin{equation}
m_{\rm eff}^2=m^2+\frac14 k^\mu(h^{-1})_{\mu\nu}k^\nu .
\end{equation}
Thus $k^\mu$ changes only the Lorentz-invariant mass parameter at
$O(k^2)$, while $\kappa^{\mu\nu}$ carries the physical Lorentz anisotropy in
Eq.~\eqref{SMEmodel}.  Earlier, various classical and quantum aspects of this theory (including calculations up to the two-loop order) have been studied in \cite{BaetaScarpelli:2003yd,Brito:2013npa,Casana:2014dfa,Casana:2014zva,Casana:2015pra,BaetaScarpelli:2021dhz,Altschul:2022isc,Brito:2023mir,Lehum:2023spy,Ageev:2025fqv,Ageev:2025bwc}.

The natural generalization of this model consists in introducing a nonminimal coupling involving dimension-5 operators, and perhaps other higher-dimensional operators (such as dimension-8 operators, see e.g. \cite{Murphy:2020rsh}), which are intensively studied in the context of elementary particle phenomenology. While, up to now, higher-dimension LV terms have been actively studied within various extensions of the spinor QED (see \cite{Kostelecky:2009zp} for motivations to studying such operators, \cite{Kostelecky:2018yfa} for the their classification, with dimensions are up to 6, and \cite{Ferrari:2018tps} for a review of perturbative impacts of such operators), no studies of higher-dimension LV operators in scalar QED were performed up to now. Within this paper, we formulate the CPT-odd scalar QED with nonminimal coupling and calculate the one-loop radiative corrections in it. Our main goal will be to calculate the one-loop contributions to the self-energy of both the scalar and gauge field by employing dimensional regularization. The main points of the computation are as follows. First, the expansion of the covariant derivative necessarily produces a mixed minimal-nonminimal seagull term proportional to $eg$, which must be kept for a complete one-loop analysis. Second, the photon two-point function should be proved to be transverse after all bubble and seagull graphs are combined. Third, the divergent part of the effective action requires both ordinary and LV counterterms. We also give the superficial degree of divergence of the higher-point functions that lie beyond the two-point scope of this calculation.

The structure of our paper reads as follows. In Section \ref{s2}, we introduce our model, specifically the CPT-odd LV scalar QED and write the Feynman rules. Section \ref{s3} is dedicated to one-loop calculations. Section \ref{s4} discusses the divergent effective action. Finally, our results are summarized in Section \ref{s5}. Throughout this work, we will employ natural units.

\section{scalar QED with vector nonminimal coupling}\label{s2}
As the starting point, let us consider the LV extended scalar QED model described by the Lagrangian (with the gauge-fixing term included):
\begin{align}\label{Lmodel}
    \mathcal{L}=-\frac{1}{4}F_{\mu\nu}F^{\mu\nu}+(\mathcal{D}_\mu\phi)^{\ast}(\mathcal{D}^\mu\phi) - m^2 \phi^{\ast}\phi-\frac{\lambda_\phi}{4}(\phi^\ast\phi)^2-\frac{1}{2\xi}(\partial_\mu A^\mu)^2,
\end{align}
where $\mathcal{D}_\mu=D_\mu+ig c_\nu \widetilde{F}^{\mu\nu}$ is the LV covariant derivative that accomplishes the nonminimal coupling, with, as usual, $D_\mu=\partial_\mu+ieA_\mu$ being the usual covariant derivative, the vector field  $B^{\mu}$ is defined as $B^{\mu}=c_\nu \widetilde{F}^{\mu\nu}$, and $\widetilde{F}^{\mu\nu}=\frac{1}{2}\epsilon^{\mu\nu\rho\sigma}F_{\rho\sigma}$ is the dual electromagnetic field tensor. Besides, $c^\nu$
represents a CPT-odd LV constant axial vector field. We note that such a coupling can be treated as a certain reminiscence of the nonminimal spinor-vector coupling used in \cite{Gomes:2009ch} to generate the CPT-even LV aether-like term in the nonminimal spinor QED. Such a vector defines a preferential direction in spacetime, which breaks the equivalence between particle and observer transformations, thereby leading to the Lorentz symmetry violation. 

The Lagrangian (\ref{Lmodel}) can be split into three parts, describing free gauge and scalar actions and the coupling term, namely
\begin{align}
 \mathcal{L}= \mathcal{L}_g + \mathcal{L}_s + \mathcal{L}_{int},    
\end{align}
where
\begin{align}
\mathcal{L}_g=-\frac{1}{4}F_{\mu\nu}F^{\mu\nu}-\frac{\lambda}{2}(\partial_\mu A^\mu)^2, 
\end{align}
\begin{align}
\mathcal{L}_s=\partial_\mu \phi^{\ast} \partial^\mu \phi - m^2 \phi^{\ast}\phi, \end{align}
and
\begin{align}
\mathcal{L}_{\mathrm{int}}=&-\ii\left(eA_\mu+gB_\mu\right)
\left(\phi^\ast\partial^\mu\phi-\phi\partial^\mu\phi^\ast\right)+e^2 A_\mu A^\mu \phi^\ast\phi\nonumber\\&+2eg A_\mu B^\mu\phi^\ast\phi
+g^2 B_\mu B^\mu\phi^\ast\phi-\frac{\lambda_\phi}{4}(\phi^\ast\phi)^2 .
\label{Lint}
\end{align}
The third term in the second line of Eq.~\eqref{Lint}, proportional to $2eg$, is required by the square of the generalized covariant derivative. It contributes to the photon two-point function through a mixed seagull graph and is essential for a complete organization of the gauge-sector divergences.

\begin{figure}[ht]
\centering
\begin{tabular}{ccc}
\begin{tikzpicture}[scale=0.9]
\draw[dashed,thick] (0,0)--(2.1,0);
\node at (1.05,0.35) {$k$};
\node at (1.05,-0.5) {$\displaystyle \frac{\ii}{k^2-m^2}$};
\end{tikzpicture}
&
\begin{tikzpicture}[scale=0.9]
\draw[decorate,decoration={snake,amplitude=1.2pt,segment length=5pt},thick] (0,0)--(2.1,0);
\node at (1.05,0.35) {$k$};
\node at (1.05,-0.5) {$\displaystyle -\frac{\ii\eta_{\mu\nu}}{k^2}$};
\end{tikzpicture}
&
\begin{tikzpicture}[scale=0.9]
\coordinate (v) at (0,0);
\draw[decorate,decoration={snake,amplitude=1.2pt,segment length=5pt},thick] (-1.2,0)--(v);
\draw[dashed,thick] (v)--(0.95,0.65);
\draw[dashed,thick] (v)--(0.95,-0.65);
\node at (0,-1.05) {$V_e^\mu$};
\end{tikzpicture}
\\[1.1cm]
\begin{tikzpicture}[scale=0.9]
\coordinate (v) at (0,0);
\draw[decorate,decoration={snake,amplitude=1.2pt,segment length=5pt},thick] (-1.2,0)--(v);
\draw[dashed,thick] (v)--(0.95,0.65);
\draw[dashed,thick] (v)--(0.95,-0.65);
\fill (v) circle (2.2pt);
\node at (0,-1.05) {$V_g^\mu$};
\end{tikzpicture}
&
\begin{tikzpicture}[scale=0.9]
\coordinate (v) at (0,0);
\draw[decorate,decoration={snake,amplitude=1.2pt,segment length=5pt},thick] (-0.95,0.65)--(v);
\draw[decorate,decoration={snake,amplitude=1.2pt,segment length=5pt},thick] (-0.95,-0.65)--(v);
\draw[dashed,thick] (v)--(0.95,0.65);
\draw[dashed,thick] (v)--(0.95,-0.65);
\node at (0,-1.05) {$V_{ee}^{\mu\nu}$};
\end{tikzpicture}
&
\begin{tikzpicture}[scale=0.9]
\coordinate (v) at (0,0);
\draw[decorate,decoration={snake,amplitude=1.2pt,segment length=5pt},thick] (-0.95,0.65)--(v);
\draw[decorate,decoration={snake,amplitude=1.2pt,segment length=5pt},thick] (-0.95,-0.65)--(v);
\draw[dashed,thick] (v)--(0.95,0.65);
\draw[dashed,thick] (v)--(0.95,-0.65);
\fill (v) circle (2.2pt);
\node at (0,-1.05) {$V_{eg}^{\mu\nu}$};
\end{tikzpicture}
\\[1.1cm]
\multicolumn{3}{c}{
\begin{tikzpicture}[scale=0.9]
\coordinate (v) at (0,0);
\draw[decorate,decoration={snake,amplitude=1.2pt,segment length=5pt},thick] (-0.95,0.65)--(v);
\draw[decorate,decoration={snake,amplitude=1.2pt,segment length=5pt},thick] (-0.95,-0.65)--(v);
\draw[dashed,thick] (v)--(0.95,0.65);
\draw[dashed,thick] (v)--(0.95,-0.65);
\fill (-0.18,0) circle (2.0pt);
\fill (0.18,0) circle (2.0pt);
\node at (0,-1.05) {$V_{gg}^{\mu\nu}$};
\end{tikzpicture}}
\end{tabular}
\caption{Vector representation of the propagators and interaction vertices of the LV scalar QED model. Wavy lines denote photons and dashed lines denote scalar fields. Filled dots indicate insertions of the LV background vector $c^\mu$. The mixed $eg$ seagull contribution and the purely nonminimal $g^2$ one are displayed explicitly; their momentum-space expressions are given by Eqs.~\eqref{props}-\eqref{4vertices}.}
\label{feynmanrules}
\end{figure}
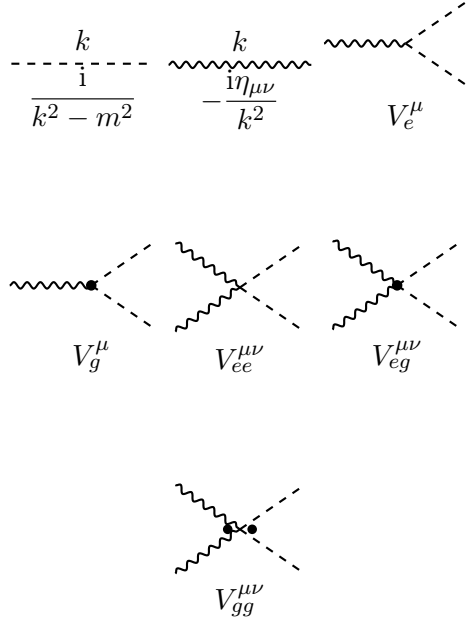

In the following calculation we use the Feynman gauge, $\xi=1$. The propagators are
\begin{equation}
    \langle A_\mu A_\nu\rangle_0(k)=\frac{-\ii\eta_{\mu\nu}}{k^2+\ii0},
    \qquad
    \langle\phi\phi^\ast\rangle_0(k)=\frac{\ii}{k^2-m^2+\ii0}.
    \label{props}
\end{equation}
For later reference, we also write the relevant vertices explicitly. With all momenta incoming, photon momentum $q$, incoming scalar momentum $p$, outgoing scalar momentum $p'$, and $q+p+p'=0$, the trilinear scalar-scalar-photon vertices are
\begin{align}
    V_e^\mu(p',p)&=\ii e(p'+p)^\mu,
    \\
    V_g^\mu(p',p;q)&=\ii g\epsilon^{\mu\kappa\alpha\beta}c_\kappa q_\alpha(p'+p)_\beta .
    \label{3vertices}
\end{align}
The two-photon scalar-scalar vertices are
\begin{align}
    V_{ee}^{\mu\nu}&=2\ii e^2\eta^{\mu\nu},
    \\
    V_{eg}^{\mu\nu}(q,r)&=2\ii eg\left(
    \epsilon^{\mu\kappa\alpha\nu}c_\kappa r_\alpha
    +\epsilon^{\nu\kappa\alpha\mu}c_\kappa q_\alpha
    \right),
    \\
    V_{gg}^{\mu\nu}(q,r)&=2\ii g^2
    \epsilon^{\rho\kappa\alpha\mu}\epsilon_{\rho}{}^{\lambda\beta\nu}
    c_\kappa c_\lambda q_\alpha r_\beta .
    \label{4vertices}
\end{align}
Here $q$ and $r$ are the photon momenta associated with the Lorentz indices $\mu$ and $\nu$, respectively. The diagrammatic representation of these rules is shown in Fig.~\ref{feynmanrules}. The four-scalar vertex following from our convention in
Eq.~\eqref{Lmodel} is
\begin{equation}
 V_{\phi^2(\phi^\ast)^2}=-\ii\lambda_\phi .
 \label{quarticvertex}
\end{equation}
It first enters the present calculation through the scalar tadpole.

\subsection{Power counting and scope}

At this point, it is useful to state explicitly what the two-point calculation does not cover. Let $E_A$ and $E_\phi$ denote the numbers of external photons and external scalar legs, and let $N_g$ count the total power of the nonminimal
coupling in a graph (a $g^2$ seagull counts twice). Each power of $g$ brings
one additional derivative relative to renormalizable scalar QED. In four
dimensions the superficial degree of divergence is therefore
\begin{equation}
 \omega=4-E_A-E_\phi+N_g .
 \label{omega}
\end{equation}
This is only a superficial bound: charge conjugation, antisymmetry of the
Levi-Civita tensor, and Ward identities can reduce $\omega$ or make a graph
vanish. At a fixed order in the EFT expansion, Eq.~\eqref{omega} gives a finite list:
\begin{center}
\begin{tabular}{c|c|c}
order & superficial degree & largest number of external legs\\ \hline
$g^0$ & $4-E_A-E_\phi$ & $4$\\
$g^1$ & $5-E_A-E_\phi$ & $5$\\
$g^2$ & $6-E_A-E_\phi$ & $6$
\end{tabular}
\end{center}
Thus, ordinary scalar QED requires, in addition to two-point
renormalization, the scalar-scalar-photon,
scalar-scalar-photon-photon, and four-scalar functions. With one insertion
of the dimension-five operator, superficially divergent five-leg functions
can occur. On the other hand, with two insertions, the bound extends to six legs. For example, $\Gamma_{\phi^\ast\phi A^n}$ has $\omega=2-n+N_g$, while a pure $n$-photon
scalar loop has $\omega=4-n+N_g$ (odd-$n$ amplitudes vanish by charge
conjugation). If arbitrary powers of $g$ are retained, there is no finite
upper bound on the number of external legs. This is the usual signal that the
model is renormalizable only order by order as an EFT. The present paper
determines the two-point counterterms through $O(g^2)$. In this sense, the higher-point counterterms listed by this power counting require separate calculations left for future works. To summarize, we have the usual gauge and scalar  propagators, whereas the LV effects come from the two new vertices of interaction. Now, with the Feynman rules at our disposal, we can proceed with studying the radiative corrections at the one-loop order for this model. 

\section{Radiative corrections}\label{s3}

This section is devoted to calculating the divergent parts
of the one-loop radiative corrections for the theoretical model presented in the previous section. For this purpose, such calculations will be performed using standard Feynman diagram approach with using of the dimensional regularization.

\subsection{Vacuum polarization}
The one-loop contribution to the photon two-point function has the following general form:
\begin{equation}
\Gamma^{(2)}_{AA}=\frac{1}{2}\int\frac{\dd^4p}{(2\pi)^4}A_\mu(-p)\Pi^{\mu\nu}(p)A_\nu(p).
\end{equation}
There is a short way to see the structure of this amplitude before doing any
integrals.  The scalar couples to the effective Abelian connection
\begin{equation}
 {\cal A}_\mu=eA_\mu+gB_\mu .
 \label{effectiveconnection}
\end{equation}
At one loop there are no internal photons in the vacuum-polarization graphs.
Consequently, the ordinary scalar-QED polarization tensor (including its
seagull) is simply evaluated with ${\cal A}_\mu$ on both external legs.
With the Fourier convention $\partial_\rho\to-\ii p_\rho$, define the
external-leg map
\begin{equation}
 {\cal K}_{\mu}{}^{\alpha}(p)=\frac{\delta{\cal A}_\mu(p)}{\delta A_\alpha(p)}
 =e\delta_\mu{}^\alpha-\ii g\,c^\nu\epsilon_{\mu\nu\rho}{}^\alpha p^\rho .
 \label{Roperator}
\end{equation}
If $\widehat\Pi^{\mu\nu}$ denotes the conventional unit-charge scalar-QED
tensor, the full result has the factorized form
\begin{equation}
 \Pi^{\alpha\beta}(p)={\cal K}_{\mu}{}^\alpha(-p)\,\widehat\Pi^{\mu\nu}(p)\,
 {\cal K}_{\nu}{}^\beta(p).
 \label{factorization}
\end{equation}
The index placement in Eqs.~\eqref{effectiveconnection}-\eqref{factorization} is important:
$p_\alpha{\cal K}_{\mu}{}^\alpha(-p)=e p_\mu$. Since
$p_\mu\widehat\Pi^{\mu\nu}=0$, Eq.~\eqref{factorization} proves
$p_\alpha\Pi^{\alpha\beta}=0$ without evaluating the loop. The explicit calculation
below is retained because it identifies the local counterterms and shows how
the bubble and seagull pieces combine.

\begin{figure}[ht]
\centering
\includegraphics[scale=0.5]{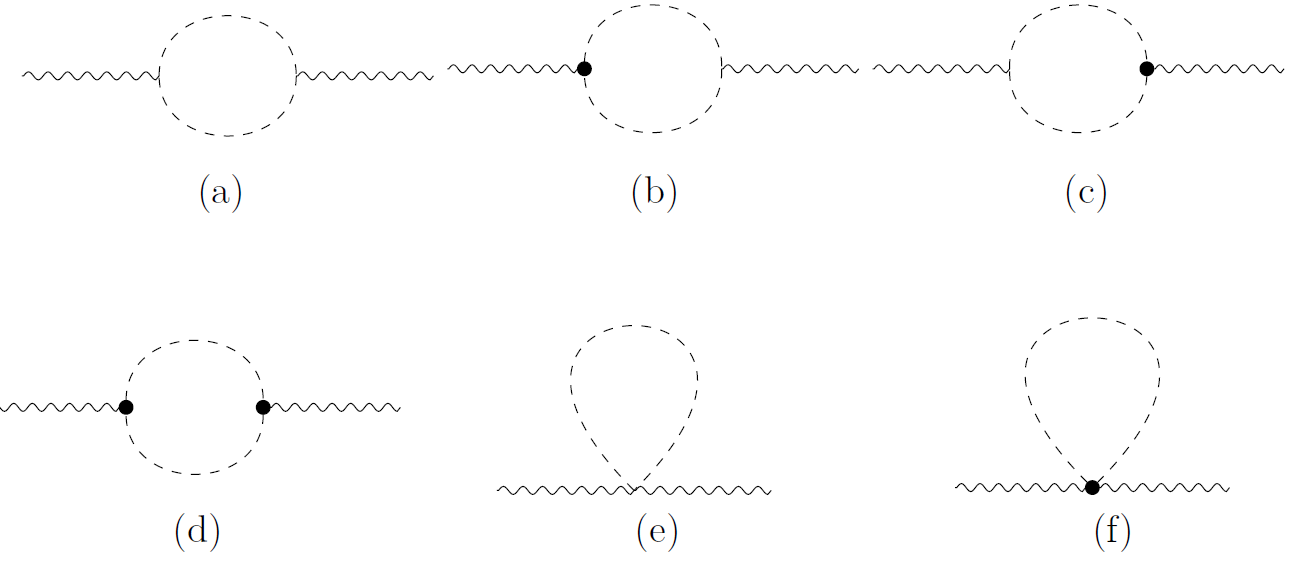}
\caption{One-loop diagrams contributing to the vacuum polarization. The complete set contains the usual scalar-QED bubble and seagull graphs, the mixed minimal-nonminimal graphs proportional to $eg$, and the purely nonminimal graphs proportional to $g^2$.}
\label{fig:vp}
\end{figure}

The full one-loop polarization tensor receives bubble contributions from pairs of trilinear vertices and seagull contributions from the quartic vertices,
\begin{equation}  \Pi^{\mu\nu}=\Pi_{ee}^{\mu\nu}+\Pi_{eg}^{\mu\nu}+\Pi_{gg}^{\mu\nu}.
\end{equation}
The relevant topologies are displayed in Fig.~\ref{fig:vp}. The mixed sector $\Pi_{eg}^{\mu\nu}$ contains both the two bubble graphs with one minimal and one nonminimal trilinear vertex and the mixed seagull graph generated by the $2egA_\mu B^\mu\phi^\ast\phi$ term. The purely nonminimal sector contains the bubble graph with two $g$ vertices and the $g^2B_\mu B^\mu\phi^\ast\phi$ seagull graph.

It is convenient to write down transverse projectors:
\begin{align}
P^{\mu\nu}(p)&=p^\mu p^\nu-p^2\eta^{\mu\nu},
\\ C^{\mu\nu}(p,c)&=\epsilon^{\mu\nu\kappa\alpha}c_\kappa p_\alpha,
\\T^{\mu\nu}(p,c)&=\eta^{\mu\nu}\left(c^2p^2-(c\cdot p)^2\right)+(c\cdot p)(c^\mu p^\nu+c^\nu p^\mu)
\nonumber\\&\hspace{1.0cm}-c^2p^\mu p^\nu-p^2c^\mu c^\nu .
\label{Ttensor}
\end{align}
They obey
\begin{equation}
    p_\mu P^{\mu\nu}=p_\mu C^{\mu\nu}=p_\mu T^{\mu\nu}=0 .
\end{equation}
Expanding the two external-leg maps in Eq.~\eqref{factorization} generates
the $e^2P^{\mu\nu}$, $egp^2C^{\mu\nu}$, and
$g^2p^2T^{\mu\nu}$ structures displayed below, with the signs and factors of
$\ii$ fixed by the Fourier and two-point-function conventions.
Within the dimensional regularization with $D=4-2\eps$, the divergent part of the usual scalar-QED contribution can be written as
\begin{equation}
\Pi^{\mu\nu}_{ee,\mathrm{div}}(p)=\frac{\ii e^2}{48\pi^2\eps}P^{\mu\nu}(p).
\label{PiUsual}
\end{equation}  
The mixed bubble contribution is proportional to $(6m^2-p^2)C^{\mu\nu}$, but the mixed $eg$ seagull graph cancels its mass-dependent part. Explicitly,
\begin{align}
\Pi^{\mu\nu}_{eg,\mathrm{bub,div}}(p)&=\frac{\ii eg}{24\pi^2\eps}(6m^2-p^2)C^{\mu\nu}(p,c),\\\Pi^{\mu\nu}_{eg,\mathrm{sg,div}}(p)
&=-\frac{\ii eg}{4\pi^2\eps}m^2 C^{\mu\nu}(p,c),
\end{align}
so that
\begin{equation}
\Pi^{\mu\nu}_{eg,\mathrm{div}}(p)=-\frac{\ii eg}{24\pi^2\eps}p^2C^{\mu\nu}(p,c).
\label{PiEG}
\end{equation}
Similarly, the mass-dependent part of the purely nonminimal bubble is canceled by the $g^2$ seagull:
\begin{align}
\Pi^{\mu\nu}_{gg,\mathrm{bub,div}}(p)&=-\frac{\ii g^2}{48\pi^2\eps}(6m^2-p^2)T^{\mu\nu}(p,c),\\\Pi^{\mu\nu}_{gg,\mathrm{sg,div}}(p)&=\frac{\ii g^2}{8\pi^2\eps}m^2T^{\mu\nu}(p,c),
\end{align}
which gives
\begin{equation}
\Pi^{\mu\nu}_{gg,\mathrm{div}}(p)=\frac{\ii g^2}{48\pi^2\eps}p^2T^{\mu\nu}(p,c).
\label{PiGG}
\end{equation}
Therefore, the complete divergent part of the photon two-point function is
\begin{equation}
\Pi^{\mu\nu}_{\mathrm{div}}(p)=\frac{\ii}{48\pi^2\eps}\left[e^2P^{\mu\nu}(p)-2egp^2C^{\mu\nu}(p,c)+g^2p^2T^{\mu\nu}(p,c)\right].
\label{Pitotal}
\end{equation}
Eq.~\eqref{Pitotal} is manifestly transverse. The cancellation of the $m^2$ pieces shows that the divergent part of the photon sector does not generate CFJ or aether counterterms from this complete set of one-loop two-point graphs. Instead, the LV photon counterterms are higher-derivative operators, as can be expected for the dimension-five nonminimal coupling  (some tree-level impacts of these terms within LV QED have been studied in \cite{Casana:2018rhg}).

\subsection{Scalar self-energy}

\begin{figure}[ht]
\centering
\includegraphics[scale=0.5]{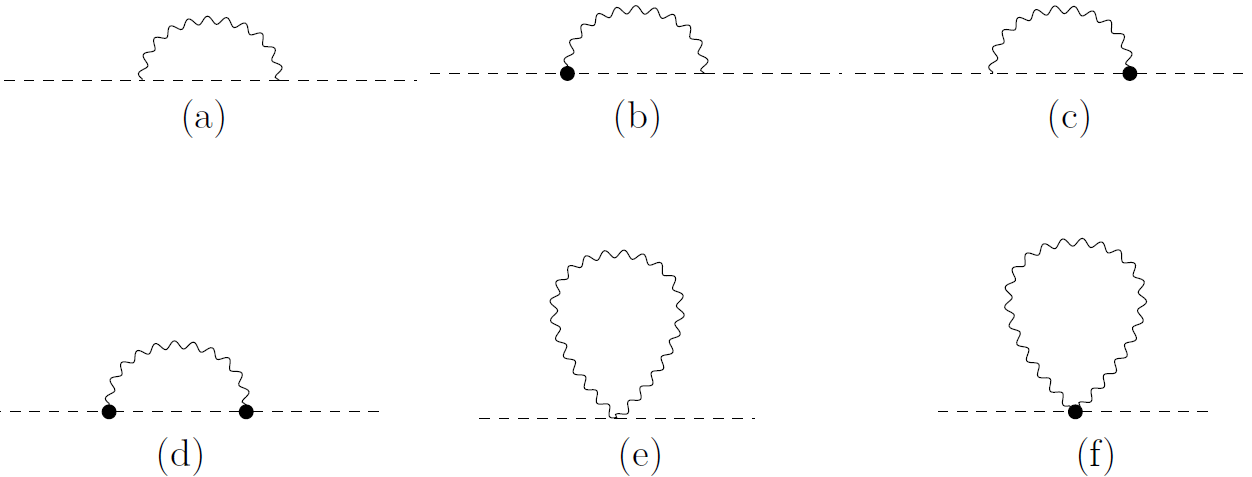}
\caption{Gauge-field diagrams contributing to the scalar self-energy. The
ordinary, mixed, and purely nonminimal vertices are denoted by zero, one, and
two filled dots. The four-scalar tadpole is included analytically in
Eq.~\eqref{SigmaLI}.}
\label{fig:self}
\end{figure}

The one-loop scalar self-energy contribution is defined by
\begin{equation}
\Gamma^{(2)}_{\phi^\ast\phi}=\int\frac{\dd^4p}{(2\pi)^4}\phi^\ast(-p)\Sigma(p)\phi(p).
\end{equation}
The gauge diagrams are shown in Fig.~\ref{fig:self}; the scalar tadpole from
Eq.~\eqref{quarticvertex} is included below. The mixed $eg$ bubble vanishes
algebraically because the Levi-Civita tensor is contracted with two identical
momentum combinations. The local photon seagulls require more care: although
their massless integrals are scaleless in pure minimal subtraction, their
ultraviolet power divergences must not be interpreted as absent. We first
quote the logarithmic poles at $D=4$ and then discuss the power-divergent
pieces separately.

The minimal gauge bubble and the quartic scalar tadpole give the
$D=4$ logarithmic pole
\begin{equation}
\Sigma_{\mathrm{LI,div}}^{(4)}(p) =\frac{\ii}{16\pi^2\eps}
\left[e^2\left(m^2+2p^2\right)+\lambda_\phi m^2\right].
\label{SigmaLI}
\end{equation}
The contributions of the first order in $g$ contain the contraction of the Levi-Civita tensor with symmetric combinations of the loop momentum and the external momentum. After the loop integration, these terms do not contribute to the divergent scalar two-point function. The leading LV scalar correction is therefore of the order $g^2$ and reads
\begin{equation}
\Sigma_{\mathrm{LV,div}}(p)=\frac{\ii g^2}{24\pi^2\eps}\left[(c\cdot p)^2-c^2p^2\right](3m^2-p^2).
\label{SigmaLV}
\end{equation}
Thus
\begin{equation}
\Sigma_{\mathrm{log,div}}^{(4)}(p)=\Sigma_{\mathrm{LI,div}}^{(4)}(p)
+\Sigma_{\mathrm{LV,div}}(p).
\end{equation}
The term proportional to $m^2$ in Eq.~\eqref{SigmaLV} is an aether-like
correction to the scalar kinetic term \cite{Gomes:2009ch}; the remaining term
contains two additional derivatives.

\subsubsection{Seagull tadpoles and power divergences}
\label{powertadpoles}

To expose the ultraviolet content of the scaleless photon tadpoles, let us define
\begin{equation}
 {\cal I}_2=\mu^{4-D}\!\int\!\frac{\dd^Dk}{(2\pi)^D}\frac{1}{k^2+\ii0},
 \qquad {\cal I}_4=\mu^{4-D}\!\int\!\frac{\dd^Dk}{(2\pi)^D}.
 \label{powerintegrals}
\end{equation}
With the symmetry factor of the closed photon line included, the local
seagulls in Feynman gauge are, in the four-dimensional tensor algebra used
here,
\begin{align}
 \Sigma_{ee,\mathrm{sg}}&=4e^2{\cal I}_2,
 & \Sigma_{eg,\mathrm{sg}}&=0,
 & \Sigma_{gg,\mathrm{sg}}&=\frac{3}{2}g^2c^2{\cal I}_4 .
 \label{seagullpower}
\end{align}
The mixed term vanishes because contraction of its two photon indices
contracts the Levi-Civita tensor with the metric. The scalar quartic tadpole
is not scaleless:
\begin{equation}
 \Sigma_{\lambda_\phi,\mathrm{sg}} =\lambda_\phi\mu^{2\eps}\int\frac{\dd^Dk}{(2\pi)^D}\frac{1}{k^2-m^2+\ii0},
 \label{scalartadpole}
\end{equation}
and its $D=4$ pole is the $\lambda_\phi m^2$ term in
Eq.~\eqref{SigmaLI}.

In conventional dimensional regularization centered at $D=4$,
${\cal I}_2={\cal I}_4=0$ because they contain no scale. This equality
combines ultraviolet and infrared information; it does not establish that
the ultraviolet mass sensitivity is zero. If the infrared part is separated
before analytic continuation, ${\cal I}_2$ contains the pole at $D=2$
associated with a quadratic divergence, while the quartic sensitivity of
${\cal I}_4$ corresponds to the $D=0$ power pole
\cite{Ossola:2003ku}. For orientation, a Lorentz-symmetric Euclidean cutoff
gives
\begin{equation}
 {\cal I}_2^{\mathrm{UV}}=\frac{\ii\Lambda^2}{16\pi^2},
 \qquad {\cal I}_4^{\mathrm{UV}}=\frac{\ii\Lambda^4}{32\pi^2},
 \label{cutoffmap}
\end{equation}
so the two photon seagulls in Eq.~\eqref{seagullpower} contain
$\ii e^2\Lambda^2/(4\pi^2)$ and
$3\ii g^2c^2\Lambda^4/(64\pi^2)$, respectively. The minimal and
nonminimal bubble graphs also contain quadratic pieces when power
divergences are retained. All such local terms are absorbed into the scalar
mass and kinetic counterterms. Their numerical coefficients are
scheme-dependent, whereas the $D=4$ logarithmic poles in
Eqs.~\eqref{SigmaLI} and \eqref{SigmaLV} are the minimal-subtraction results
used in the rest of the paper.

\section{One-loop divergences and counterterms}\label{s4}

The divergent part of the one-loop two-point effective action is
\begin{align}
\Gamma_{\mathrm{div}}^{(2)}=&\frac{1}{2}\int\frac{\dd^4p}{(2\pi)^4}A_\mu(-p)\Pi^{\mu\nu}_{\mathrm{div}}(p)A_\nu(p)\nonumber\\&+\int\frac{\dd^4p}{(2\pi)^4}
\phi^\ast(-p)\Sigma_{\mathrm{log,div}}^{(4)}(p)\phi(p).
\label{Gammadiv}
\end{align}
Equation~\eqref{Gammadiv} identifies the logarithmic two-point counterterms.
The ordinary scalar-QED pieces renormalize the Maxwell term, the scalar
kinetic and mass terms, and the mass-dependent part of the scalar quartic
tadpole. In addition, Eq.~\eqref{seagullpower} requires a local scalar-mass
counterterm containing the quadratic and quartic power sensitivities. In a
Wilsonian notation,
\begin{equation}
 {\cal L}_{\rm ct}\supset-\delta m_{\rm power}^2\,\phi^\ast\phi,\qquad
 \delta m_{\rm power}^2=O(e^2\Lambda^2,\lambda_\phi\Lambda^2,g^2c^2\Lambda^4).
 \label{masspowerct}
\end{equation}
The precise coefficients depend on the regulator and subtraction
prescription.

The mixed $eg$ photon contribution is proportional to $p^2C^{\mu\nu}$; hence the lower-derivative CFJ divergence cancels after the mixed seagull graph is included. The corresponding gauge-sector counterterm is the higher-derivative CFJ-like operator
\begin{equation}
\mathcal{O}_{\mathrm{hCFJ}}=c_\kappa\epsilon^{\kappa\lambda\mu\nu}A_\lambda\Box F_{\mu\nu}.
\label{hCFJop}
\end{equation}
This operator is gauge invariant up to a surface term, in the same sense as the usual CFJ structure. We note that the higher-derivative CFJ-like term also arises in spinor LV QED, see e.g. \cite{Mariz:2010fm} (however, unlike our theory, the result in \cite{Mariz:2010fm} is finite).

The purely nonminimal $g^2$ photon contribution is proportional to $p^2T^{\mu\nu}$. The ordinary lower-derivative aether divergence cancels between the bubble and seagull pieces, and the required divergent gauge counterterm may be written as
\begin{equation}
\mathcal{O}_{\mathrm{hA}}=(c_\mu F^{\mu\nu})\Box(c^\lambda F_{\lambda\nu}),
\label{haetherop}
\end{equation}
up to integrations by parts and equivalent rearrangements of derivatives. In momentum space Eq.~\eqref{haetherop} reproduces the higher-derivative aether-like $p^2T^{\mu\nu}$ structure in Eq.~\eqref{PiGG}.

The two-point scalar divergence admits the gauge-covariant completions
\begin{equation}
\mathcal{O}_{\phi,1}=(c^\mu c^\nu-c^2\eta^{\mu\nu})(D_\mu\phi)^\ast(D_\nu\phi),
\label{scalarop1}
\end{equation}
and
\begin{equation}
\mathcal{O}_{\phi,2}=(c^\mu c^\nu-c^2\eta^{\mu\nu})\frac12\left[(D_\mu\phi)^\ast D^2(D_\nu\phi)+\mathrm{h.c.}\right],
\label{scalarop2}
\end{equation}
which reduce at quadratic order to the $3m^2$ and $-p^2$ structures in
Eq.~\eqref{SigmaLV}. Covariantizing these terms also produces three- and
higher-point vertices. Commutators of covariant derivatives generate
additional $F_{\mu\nu}$ operators, so the two-point calculation alone cannot
fix a unique complete higher-point basis.

The structures in Eqs.~\eqref{hCFJop}-\eqref{scalarop2} are the subset fixed
by the present two-point functions. A complete one-loop EFT treatment also
requires the charge and seagull vertex counterterms, the quartic counterterm
$\delta\lambda_\phi$, and the five- and six-leg counterterms allowed by
Eq.~\eqref{omega} at $O(g)$ and $O(g^2)$. We therefore make no claim that the
two-point calculation supplies the complete one-loop operator basis.

\section{Final remarks}\label{s5}

In this work, we have analyzed the divergent one-loop two-point functions of a LV scalar-QED model containing the CPT-odd nonminimal coupling proportional to $g c^\nu\widetilde F_{\mu\nu}$. A careful expansion of the generalized covariant derivative shows that the interaction Lagrangian contains the usual minimal scalar-QED vertices, the nonminimal trilinear vertex, the purely nonminimal seagull vertex, and a mixed minimal-nonminimal seagull vertex proportional to $eg$. Including all these terms is essential for the gauge invariance of the Lagrangian.

The vacuum polarization contains the ordinary scalar-QED divergence, a mixed CPT-odd contribution proportional to $eg$, and a CPT-even contribution proportional to $g^2$. After the mixed and purely nonminimal seagull graphs are included, the $m^2$ pieces in the LV photon sector cancel. The final divergent photon self-energy is
\begin{equation}
    \Pi^{\mu\nu}_{\mathrm{div}}(p)
    =\frac{\ii}{48\pi^2\eps}
    \left[e^2P^{\mu\nu}-2egp^2C^{\mu\nu}+g^2p^2T^{\mu\nu}\right],
\end{equation}
which is transverse, $p_\mu\Pi^{\mu\nu}_{\mathrm{div}}=0$. We note that the LV divergent terms in this expression are only higher-derivative ones, there is no divergent correction either to CFJ or to the aether term. Gauge invariance is therefore preserved by the regularized one-loop result. The scalar self-energy receives the usual scalar-QED divergence and the LV contribution proportional to $g^2[(c\cdot p)^2-c^2p^2](3m^2-p^2)$, so, we have the aether-like and higher-derivative LV corrections in the scalar sector. The terms linear in $g$ vanish in the divergent scalar two-point function by symmetric properties of the corresponding integral over momenta.

The divergent effective action reveals the counterterm structure required by the model: the Maxwell and scalar wave-function counterterms, a higher-derivative CFJ-type term, a higher-derivative CPT-even aether-like gauge operator, and LV scalar kinetic and higher-derivative operators. Since the coupling $g$ has negative mass dimension when $c^\mu$ is dimensionless, these results should be interpreted in the sense of effective field theory. The one-loop calculation identifies the operator basis that must be retained for a consistent perturbative treatment of this CPT-odd LV scalar electrodynamics. We note that there is no divergent correction matching the form of the usual CFJ term, which can be treated as a reminiscence of \cite{Lehum:2023spy}, where the CFJ term in the LV scalar QED was shown to be finite.

The possible continuation of this study could consist in application of this result within phenomenological context, for example, in studies of the Higgs mechanism in LV QED. We plan to perform this study in one of the forthcoming papers.

\section*{Funding and/or Conflicts of interests/Competing interests}
\hspace{0.5cm} The authors thank the  Coordena\c{c}\~{a}o de Aperfei\c{c}oamento de Pessoal de N\'{i}vel Superior (CAPES), Paraiba State Research Foundation (FAPESQ-PB) and the Conselho Nacional de Desenvolvimento Cient\'{i}fico e Tecnol\'{o}gico (CNPq). The author Fernando Belchior would like to to express gratitude to the Conselho Nacional de Desenvolvimento Cient\'{i}fico e Tecnol\'{o}gico CNPq for grant No. 151845/2025-5.  Albert Yu. Petrov thanks the Brazilian agency FAPESQ-PB (process No. 150891/2023-7) and CNPq (grant No. 303777/2023-0).  Paulo J. Porf\'irio thanks the Brazilian agency FAPESQ-PB (process No. 150891/2023-7) and CNPq (grant No. 307628/2022-1).

{\bf Conflicts of interests:} The authors declare no competing interests.

{\bf Data availability statement:} No data were generated or studied within this paper.

\appendix

\section{Dimensional regularization conventions}
\label{dimreg}

We use dimensional regularization by defining $D=4-2\eps$ and performing the replacement
\begin{equation}
\int\frac{\dd^4k}{(2\pi)^4}\to \mu^{2\eps}\int\frac{\dd^Dk}{(2\pi)^D}.
\end{equation}
We can write a basic scalar integral as follows
\begin{equation}
J_n(\Delta)=\mu^{2\eps}\int\frac{\dd^D\ell}{(2\pi)^D}\frac{1}{(\ell^2-\Delta+\ii0)^n}
=\frac{\ii(-1)^n}{(4\pi)^{D/2}}\frac{\Gamma(n-D/2)}{\Gamma(n)}
(\Delta)^{D/2-n}.
\end{equation}
The logarithmic $D=4$ poles used in the minimal-subtraction calculation are
\begin{align}
J_1(\Delta)&=\frac{\ii\Delta}{16\pi^2\eps}+\mathrm{finite},
\\
J_2(\Delta)&=\frac{\ii}{16\pi^2\eps}+\mathrm{finite},
\\
\mu^{2\eps}\int\frac{\dd^D\ell}{(2\pi)^D}
\frac{\ell^2}{(\ell^2-\Delta)^2}
&=\frac{2\ii\Delta}{16\pi^2\eps}+\mathrm{finite}.
\end{align}
Equations above do not by themselves dispose of power divergences. A
quadratically divergent one-loop integral appears as a UV pole at $D=2$ when
the integral is analytically continued from its convergence domain; a
quartic divergence is associated with $D=0$. A scaleless integral is zero in
pure dimensional regularization only after its UV and IR pieces have been
combined. Section~\ref{powertadpoles} keeps the corresponding local
counterterms separate from the $D=4$ logarithmic poles.

Tensor integrals are reduced according to symmetric integration,
\begin{align}
    \int\frac{\dd^D\ell}{(2\pi)^D}\ell_\mu f(\ell^2)&=0,
    \\
    \int\frac{\dd^D\ell}{(2\pi)^D}\ell_\mu\ell_\nu f(\ell^2)
    &=\frac{\eta_{\mu\nu}}{D}\int\frac{\dd^D\ell}{(2\pi)^D}\ell^2 f(\ell^2),
    \\
    \int\frac{\dd^D\ell}{(2\pi)^D}\ell_\mu\ell_\nu\ell_\rho\ell_\sigma f(\ell^2)
    &=\frac{\eta_{\mu\nu}\eta_{\rho\sigma}+\eta_{\mu\rho}\eta_{\nu\sigma}+\eta_{\mu\sigma}\eta_{\nu\rho}}{D(D+2)}
    \int\frac{\dd^D\ell}{(2\pi)^D}(\ell^2)^2 f(\ell^2).
\end{align}
For two massive scalar denominators,
\begin{equation}
D_k=k^2-m^2+\ii0, \qquad D_{k+p}=(k+p)^2-m^2+\ii0,
\end{equation}
we combine them as
\begin{equation}
\frac{1}{D_kD_{k+p}}=\int_0^1\dd x\, \frac{1}{[(k+xp)^2-\Delta_s+\ii0]^2},
\qquad \Delta_s=m^2-x(1-x)p^2,
\end{equation}
and shift $k\to\ell-xp$. For one photon and one scalar denominator in the scalar self-energy,
\begin{equation}
L_k=k^2+\ii0, \qquad D_{p-k}=(p-k)^2-m^2+\ii0,
\end{equation}
we use
\begin{equation}
\frac{1}{L_kD_{p-k}}=\int_0^1\dd x\, \frac{1}{[(k-(1-x)p)^2-\Delta_\phi+\ii0]^2},
\qquad \Delta_\phi=(1-x)m^2-x(1-x)p^2.
\end{equation}
The Levi-Civita tensor is kept as a four-dimensional object in the external tensor algebra. This prescription is sufficient for the present one-loop two-point functions because no ambiguous Dirac trace involving $\gamma_5$ occurs. All loop momenta are dimensionally continued, while contractions involving $\epsilon^{\mu\nu\alpha\beta}$ are performed after all loop integrals are simplified, and the space-time dimension is taken to be equal to 4, so we have a kind of dimensional reduction.

\section{One-loop integrals for the vacuum polarization}\label{vpints}

In this appendix we list the explicit integrals used to obtain Eq.~\eqref{Pitotal}. We define
\begin{equation}
K^\mu=2k^\mu+p^\mu, \qquad G^\mu(k,p)=\epsilon^{\mu\kappa\alpha\beta}c_\kappa p_\alpha K_\beta,
\end{equation}
where $p$ is the external photon momentum. The bubble integrals are
\begin{align}
\Pi^{\mu\nu}_{ee,\mathrm{bub}}(p) &=e^2\mu^{2\eps}\int\frac{\dd^Dk}{(2\pi)^D}
\frac{K^\mu K^\nu}{D_kD_{k+p}},
\label{PeeBubbleInt}
\\ \Pi^{\mu\nu}_{eg,\mathrm{bub}}(p)
&=eg\mu^{2\eps}\int\frac{\dd^Dk}{(2\pi)^D}
\frac{K^\mu G^\nu(k,p)+G^\mu(k,p)K^\nu}{D_kD_{k+p}},
\label{PegBubbleInt}
\\ \Pi^{\mu\nu}_{gg,\mathrm{bub}}(p) &=g^2\mu^{2\eps}\int\frac{\dd^Dk}{(2\pi)^D}
\frac{G^\mu(k,p)G^\nu(k,p)}{D_kD_{k+p}}.
\label{PggBubbleInt}
\end{align}
The seagull integrals are
\begin{align}
    \Pi^{\mu\nu}_{ee,\mathrm{sg}}(p)
    &=-2e^2\eta^{\mu\nu}\mu^{2\eps}\int\frac{\dd^Dk}{(2\pi)^D}\frac{1}{D_k},
    \label{PeeSeagullInt}
    \\
    \Pi^{\mu\nu}_{eg,\mathrm{sg}}(p)
    &=-\mu^{2\eps}\int\frac{\dd^Dk}{(2\pi)^D}
    \frac{V_{eg}^{\mu\nu}(-p,p)}{\ii D_k},
    \label{PegSeagullInt}
    \\
    \Pi^{\mu\nu}_{gg,\mathrm{sg}}(p)
    &=-\mu^{2\eps}\int\frac{\dd^Dk}{(2\pi)^D}
    \frac{V_{gg}^{\mu\nu}(-p,p)}{\ii D_k}.
    \label{PggSeagullInt}
\end{align}
Here the factors of $\ii$ are arranged so that the final two-point function is defined by $\Gamma^{(2)}_{AA}=\frac12 A_{\mu}\Pi^{\mu\nu} A_{\nu}$, consistently with the Feynman rules of Eqs.~\eqref{3vertices} and \eqref{4vertices}. Substituting the explicit seagull vertices gives
\begin{align}
    \Pi^{\mu\nu}_{eg,\mathrm{sg}}(p)
    &=-4egC^{\mu\nu}(p,c)\mu^{2\eps}\int\frac{\dd^Dk}{(2\pi)^D}\frac{1}{D_k},
    \\
    \Pi^{\mu\nu}_{gg,\mathrm{sg}}(p)
    &=2g^2T^{\mu\nu}(p,c)\mu^{2\eps}\int\frac{\dd^Dk}{(2\pi)^D}\frac{1}{D_k},
\end{align}
with the sign and normalization following from the momentum assignment $(-p,p)$ in the two external photon legs.

After Feynman parametrization and the shift $k\to\ell-xp$, the divergent parts of the bubble graphs are written as
\begin{align}
    \Pi^{\mu\nu}_{ee,\mathrm{bub,div}}(p)
    &=\frac{\ii e^2}{16\pi^2\eps}
    \left[\frac{1}{3}p^\mu p^\nu+\left(2m^2-\frac{p^2}{3}\right)\eta^{\mu\nu}\right],
    \\
    \Pi^{\mu\nu}_{ee,\mathrm{sg,div}}(p)
    &=-\frac{\ii e^2m^2}{8\pi^2\eps}\eta^{\mu\nu},
\end{align}
which combine to Eq.~\eqref{PiUsual}. In the LV sector one obtains
\begin{align}
    \Pi^{\mu\nu}_{eg,\mathrm{bub,div}}(p)
    &=\frac{\ii eg}{24\pi^2\eps}(6m^2-p^2)C^{\mu\nu}(p,c),
    \\
    \Pi^{\mu\nu}_{eg,\mathrm{sg,div}}(p)
    &=-\frac{\ii eg}{4\pi^2\eps}m^2C^{\mu\nu}(p,c),
    \\
    \Pi^{\mu\nu}_{gg,\mathrm{bub,div}}(p)
    &=-\frac{\ii g^2}{48\pi^2\eps}(6m^2-p^2)T^{\mu\nu}(p,c),
    \\
    \Pi^{\mu\nu}_{gg,\mathrm{sg,div}}(p)
    &=\frac{\ii g^2}{8\pi^2\eps}m^2T^{\mu\nu}(p,c).
\end{align}
The $m^2$ terms cancel in the sums $\Pi_{eg,\mathrm{bub}}+\Pi_{eg,\mathrm{sg}}$ and $\Pi_{gg,\mathrm{bub}}+\Pi_{gg,\mathrm{sg}}$, yielding Eqs.~\eqref{PiEG} and \eqref{PiGG}.

\section{One-loop integrals for the scalar self-energy}\label{selfints}

For the scalar self-energy we use an internal photon momentum $k$ and define
\begin{equation}
R^\mu=2p^\mu-k^\mu, \qquad H^\mu(k,p)=\epsilon^{\mu\kappa\alpha\beta}c_\kappa k_\alpha R_\beta,
\end{equation}
where $p$ is the external scalar momentum. The ordinary scalar-QED bubble integral is
\begin{equation}
\Sigma_{ee,\mathrm{bub}}(p)=e^2\mu^{2\eps}\int\frac{\dd^Dk}{(2\pi)^D}
\frac{R^2}{L_kD_{p-k}}.
\label{SeeBubbleInt}
\end{equation}
Using the Feynman parameter formula in Appendix~\ref{dimreg}, shifting $k\to\ell+(1-x)p$, and dropping odd powers of $\ell$, one obtains
\begin{align}
\Sigma_{ee,\mathrm{bub,div}}(p)&=\frac{\ii e^2}{16\pi^2\eps}\int_0^1\dd x\left[2\Delta_\phi+(1+x)^2p^2\right]\\&=\frac{\ii e^2}{16\pi^2\eps}(m^2+2p^2).
\end{align}
The ordinary photon seagull is
\begin{equation}
    \Sigma_{ee,\mathrm{sg}}(p)=4e^2{\cal I}_2.
\end{equation}
It has no $D=4$ logarithmic pole in pure minimal subtraction, but its
$D=2$ ultraviolet pole contributes to the power-sensitive mass counterterm.
The scalar quartic interaction gives
\begin{equation}
 \Sigma_{\lambda_\phi,\mathrm{sg}}(p)=\lambda_\phi J_1(m^2)=\frac{\ii\lambda_\phi m^2}{16\pi^2\eps}+\mathrm{finite},
\end{equation}
which completes Eq.~\eqref{SigmaLI}.

The mixed minimal-nonminimal bubble is
\begin{equation}
\Sigma_{eg,\mathrm{bub}}(p)=2eg\mu^{2\eps}\int\frac{\dd^Dk}{(2\pi)^D}
\frac{R_\mu H^\mu(k,p)}{L_kD_{p-k}}.
\label{SegBubbleInt}
\end{equation}
However, $R_\mu H^\mu=\epsilon^{\mu\kappa\alpha\beta}R_\mu c_\kappa k_\alpha R_\beta=0$, because the Levi-Civita tensor is contracted with two identical vectors $R_\mu$ and $R_\beta$. Thus
\begin{equation}
    \Sigma_{eg,\mathrm{bub}}(p)=0.
\end{equation}
The mixed seagull tadpole vanishes after contraction of the two photon
indices,
\begin{equation}
    \Sigma_{eg,\mathrm{sg}}(p)=0.
\end{equation}
The purely nonminimal bubble is
\begin{equation}
\Sigma_{gg,\mathrm{bub}}(p)=g^2\mu^{2\eps}\int\frac{\dd^Dk}{(2\pi)^D}\frac{H_\mu(k,p)H^\mu(k,p)}{L_kD_{p-k}}.
\label{SggBubbleInt}
\end{equation}
After the Feynman-parameter shift and tensor reduction, its divergent part is
\begin{equation}
\Sigma_{gg,\mathrm{bub,div}}(p)=\frac{\ii g^2}{24\pi^2\eps}
\left[(c\cdot p)^2-c^2p^2\right](3m^2-p^2).
\end{equation}
The purely nonminimal seagull is
\begin{equation}
    \Sigma_{gg,\mathrm{sg}}(p)=\frac32 g^2c^2{\cal I}_4 .
\end{equation}
It has no logarithmic pole at $D=4$, so Eq.~\eqref{SigmaLV} follows from the
bubble integral \eqref{SggBubbleInt} in minimal subtraction. Nevertheless,
its quartic ultraviolet sensitivity contributes to
Eq.~\eqref{masspowerct}, as discussed in Sec.~\ref{powertadpoles}.

\section{Transversality checks}\label{transverse}

The final divergent photon self-energy is expressed in terms of the three tensors $P^{\mu\nu}$, $C^{\mu\nu}$, and $T^{\mu\nu}$ defined in Eq.~\eqref{Ttensor}. The first two contractions are immediate:
\begin{equation}
p_\mu P^{\mu\nu}=p^2p^\nu-p^2p^\nu=0, \qquad p_\mu C^{\mu\nu}=p_\mu\epsilon^{\mu\nu\kappa\alpha}c_\kappa p_\alpha=0.
\end{equation}
The last equality follows from antisymmetry in two powers of $p$. For the CPT-even tensor one obtains
\begin{align}
p_\mu T^{\mu\nu}=&p^\nu\left(c^2p^2-(c\cdot p)^2\right)+(c\cdot p)^2p^\nu+(c\cdot p)c^\nu p^2
\nonumber\\
&-c^2p^2p^\nu-p^2(c\cdot p)c^\nu=0.
\end{align}
Therefore
\begin{equation}
    p_\mu\Pi^{\mu\nu}_{\mathrm{div}}(p)=0,
\end{equation}
which proves the one-loop Ward identity for the photon two-point function.

\end{document}